\documentclass{elsart}

\usepackage{graphicx}
\usepackage{amssymb}

\begin{document}

\begin{frontmatter}

\title{Dynamical systems with time-dependent coupling: Clustering and critical
behaviour}

\author[M,Z]{Dami\'an H. Zanette}
\ead{zanette@cab.cnea.gov.ar}
\and
\author[M]{Alexander S. Mikhailov}
\ead{mikhailov@fhi-berlin.mpg.de}

\address[M]{Fritz Haber Institut der Max Planck Gesellschaft, Abteilung
Physikalische Chemie, Faradayweg 4-6, 14195 Berlin, Germany}

\address[Z]{Consejo Nacional de Investigaciones Cient\'{\i}ficas y
T\'ecnicas, Centro At\'omico Bariloche and Instituto Balseiro,
8400 Bariloche, R\'{\i}o Negro, Argentina}

\begin{abstract}
We study the collective behaviour of an ensemble of coupled
motile elements whose interactions depend on time and are
alternatively attractive or repulsive. The evolution of
interactions is driven by individual internal variables with
autonomous dynamics. The system exhibits different dynamical
regimes, with various forms of collective organization,
controlled by the range of interactions and the dispersion of
time scales in the evolution of the internal variables. In the
limit of large interaction ranges, it reduces to an ensemble of
coupled identical phase oscillators and, to some extent, admits
to be treated analytically. We find and characterize a transition
between ordered and disordered states, mediated by a regime of
dynamical clustering.
\end{abstract}

\end{frontmatter}

\section{Introduction}

Large ensembles of coupled dynamical systems are a basic tool for
the study of emergent collective behaviour in systems of
interacting agents \cite{Mikh}. Self-organization in spatially
distributed ensembles is revealed by the appearance of
spatiotemporal structures, which may exhibit complex evolution.
Instead, coupled ensembles for which space is not relevant to the
dynamical laws become organized in temporal patterns, typically,
in some form of synchronous evolution. Much attention has been
focused on synchronization phenomena over the last years
\cite{bocca1}. They have been identified in a broad class of
natural and artificial systems, and many of them have been
successfully explained by means of relatively simple models.

The degree of collective organization in a coupled ensemble,
measured by the correlation between individual motions, varies
considerably between different kinds of synchronization patterns.
Most studies on these phenomena have dealt with patterns of
highly correlated evolution, such as full and phase
synchronization. Indeed, these forms of synchronization are
essential to the functioning of some artificial systems, and have
been observed in certain insect populations \cite{ff,crick}. On
the other hand, they are expected to play a less relevant role in
most biological systems, where the complexity of collective
functions requires a delicate balance between coherence and
diversity \cite{Mikh}. Consider, for instance, the brain, where
highly coherent activity patterns are only realized under
pathological states, such as during epileptic seizures. Many
biological systems consisting of interacting agents, ranging from
biomolecular complexes to social populations, are normally found
in configurations where the ensemble is segregated into groups
with specific functions. While the evolution of individual
elements is highly correlated inside each group, the collective
dynamics of different groups is much more independent. Clustering
has been modeled by means of coupled ensembles with specific
individual dynamics or interactions
\cite{kcl,zmr,zmn,golomb,okuda}.

Usually, clustering is a dynamical process, where groups may
preserve their identity in spite of the fact that single elements
are continuously migrating between them. Intermittent formation
of clusters and migration of elements has been observed under the
action of noise \cite{mato,noise,kori}. The individual motion
towards or away from clusters may also be controlled by the
internal state of each element, which favors or inhibits grouping
with other elements. This is observed in natural phenomena ranging
from complex chemical reactions, where biomolecules react with
each other only when they have reached appropriate internal
configurations \cite{stange}, to social systems, where the
appearance of organizational structures requires compatibility
between the individual changing states of the involved agents
\cite{axel}. Focusing on this motivation, in this paper we study
a model of coupled motile elements where interactions depend on
the internal state of each element. Individual internal variables
change with time according to a prescribed autonomous evolution
law, so that interactions are also time-dependent, and result to
be alternatively attractive or repulsive. This intricate
interaction pattern induces several complex regimes of collective
evolution, including dynamical clustering. Clustering in systems
of interacting moving particles with internal dynamics has been
demonstrated when the internal states of different elements are
mutually coupled and their individual evolution is chaotic
\cite{shk}. In such case, the synchronization of internal
variables results into spatial organization, through the effect
of those variables on the degrees of freedom associated with
motion. Here, we show that similar forms of collective behaviour
are already possible with simple, autonomous oscillatory dynamics
for the internal variables.

In the next section, we introduce the model and present numerical
results for the case of two-dimensional motion in a bounded
domain. The occurrence of different dynamical regimes is
controlled by the range of interactions and by the dispersion in
the time scales associated with the internal dynamics. In Section
\ref{sect2}, the limit of large interaction ranges is analyzed.
In this limit, the system reduces to an ensemble of identical
phase oscillators with internal states. As the dispersion in the
time scales of internal variables is varied, it exhibits an
order-disorder transition between a configuration where the phase
and the internal variable are highly correlated and a uncorrelated
state. This critical transition, which we are able to describe
analytically, is mediated by a regime of dynamical clustering.
Both the transition and the clustering regime are characterized
by means of suitably defined order parameters.  In the final
section, we discuss our results with emphasis in the features
that might be also observed  in other systems of similar kinds.

\section{Dynamical clustering of motile elements}
\label{sect1}

\subsection{Model}

We consider an ensemble of $N$ motile elements at positions ${\bf
r}_i$ ($i=1,\dots,N$). The elements interact through pair forces
${\bf F}({\bf r}_i-{\bf r}_j)$ depending on their relative
positions. Moreover, the interaction of each pair is modulated by
a function $V(\theta_i,\theta_j)$ of the variables $\theta_i$ and
$\theta_j$, which characterize the internal state of the
respective elements. This modulation aims at describing the
effects of the internal states on the spatial dynamics. The
motion of the elements is assumed to be overdamped, so that the
equation of motion for element $i$ is
\begin{equation} \label{eqmot1}
\dot {\bf r}_i = \sum_{j\neq i} V(\theta_i,\theta_j) {\bf F}({\bf
r}_i-{\bf r}_j) .
\end{equation}

From the viewpoint of the numerical resolution of
Eqs.~(\ref{eqmot1}), a convenient choice for the interactions a
truncated elastic force,
\begin{equation} \label{ff}
{\bf F}({\bf r}_i-{\bf r}_j) = \left\{
\begin{array}{ll}
-k({\bf r}_i-{\bf r}_j) & \mbox{  if $|{\bf r}_i-{\bf r}_j|<r_0$,} \\
0 & \mbox{  otherwise.}
\end{array} \right.
\end{equation}
Here, $k>0$ is the elastic constant and $r_0$ is the interaction
range.

As for the internal state described by the variable $\theta_i$,
we assume that each element behaves as an autonomous oscillator
with constant frequency $\Omega_i$, much like a ``biological
clock.'' Biological periodic phenomena at several levels, from
intracellular chemical reactions to complex long-period rhythms,
are in fact well described by such elementary cyclic processes
\cite{Winfree}. The internal variable represents thus a phase,
and evolves in time as
\begin{equation} \label{eqtet}
\theta_i(t) = \Omega_it+\theta_i(0).
\end{equation}
To take into account inhomogeneity in the ensemble, the
frequencies $\Omega_i$ are drawn from a given distribution
$g(\Omega)$. Here, we take the Gaussian
\begin{equation} \label{Gaussian}
g(\Omega)= \frac{1}{\sqrt{2\pi \sigma^2}} \exp\left[
-\frac{(\Omega-\Omega_0)^2}{2\sigma^2}\right].
\end{equation}
The autonomous dynamics  of the internal phases $\theta_i$
represents cyclic processes which, in our model, are not
influenced by external factors but affect the interaction between
elements. Specifically, the interaction weight
$V(\theta_i,\theta_j)$ in Eq.~(\ref{eqmot1}) is chosen as
\begin{equation} \label{vv}
V(\theta_i,\theta_j) = \cos (\theta_i-\theta_j) =
\cos[(\Omega_i-\Omega_j)t +   \theta_{ij}^0],
\end{equation}
with $ \theta_{ij}^0=\theta_i(0)-\theta_j(0)$. With this choice,
the interaction between elements $i$ and $j$ changes its sign
periodically, in a time scale of order $(\Omega_i -
\Omega_j)^{-1}$. Thus, it is alternatively attractive or
repulsive, respectively favoring or inhibiting the aggregation of
elements. At the level of the whole ensemble, this variations give
rise to a complex time-dependent interaction pattern which, as
shown below, induces a variety of dynamical regimes, including
dynamical clustering. Note that, according to Eq.~(\ref{vv}),
interactions between pairs of elements do not depend separately
on the individual frequencies, but on the frequency differences
$\Omega_i-\Omega_j$. Thus, in Eq.~(\ref{Gaussian}), we can fix
$\Omega_0=0$.

As a final ingredient of the model, it is necessary to consider
that the ensemble is confined within a bounded spatial domain. In
fact, when mutually attracting elements have aggregated into
clusters, we expect that  the remnant repulsion between elements
in different neighbor clusters drives them away from each other.
In the long run, this effect would lead to the total dispersion of
the ensemble. Therefore, we assume that the system evolves into a
closed domain with reflecting boundaries.

\subsection{Numerical results}

We have extensively studied the above model, by solving
Eqs.~(\ref{eqmot1}) numerically for a two-dimensional system,
where an ensemble of $N=100$ elements is confined within a circle
of radius $r_{\max}$.  Simultaneous rescaling of frequencies and
time makes it possible to fix the elastic constant to $k=1$.
Moreover, we choose $r_{\max}$ as the unit of distance. In this
way, the only relevant parameters in the model are the
interaction range $r_0$ and the frequency dispersion $\sigma$.

Truncation errors in the numerical resolution of the equations of
motion (\ref{eqmot1}) may cause that, under the effect of
attractive interactions, two elements collapse to numerically
indistinguishable positions. This is expected to happen, in
particular, when the evolution of internal phases is slow (small
$\sigma$) so that attractive interactions can act during
sufficiently long times. Under such conditions, in view of
Eq.~(\ref{ff}), the mutual interaction of the collapsed elements
will vanish, and the forces applied on each of them by any other
element will be identical. Consequently, the collapsed elements
will remain ``stuck together'' for the remaining of the
evolution. To avoid this numerical artifact, we introduce in
Eqs.~(\ref{eqmot1}) a small additive-noise term which prevents
the spatial collapse of elements and, otherwise, does not affect
the dynamics.

\begin{figure}
\centering
\resizebox{.7\columnwidth}{!}{\includegraphics[clip=]{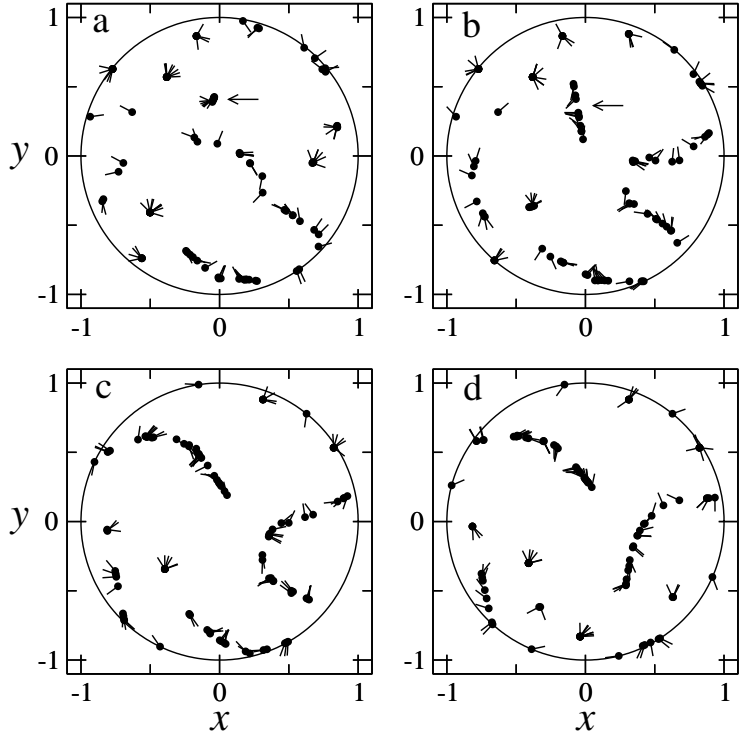}}
\caption{Four snapshots of an ensemble of size $N=100$, with
$r_0=0.3$ and $\sigma=0.1$, at times (a) $t=5100$, (b) $t=5110$,
(c) $t=5120$, and (d) $t=5130$. Each dot represents the position
${\bf r}_i=(x_i,y_i)$ of an element on the plane, while the
orientation of the stick starting at each dot indicates the value
of the corresponding internal phase $\theta_i$. Circles show the
system boundary. All the elements of the cluster marked with an
arrow in (a) are part of the ``string'' marked in (b).
\label{figr2}}
\end{figure}

Figure \ref{figr2} shows four successive snapshots of the
ensemble, in a single realization with $r_0=0.3$ and
$\sigma=0.1$. We find that many elements are entrained in several
compact clusters, while others are distributed in space along
curved lines, forming ``strings.''   These ``strings'' result
from the instabilization of clusters, as illustrated in
Figs.~\ref{figr2}a and b. Generally,  the distribution of
internal phases $\theta_i$ of the elements belonging a given
cluster spans a total angle less than $\pi$, but this total angle
is often larger than $\pi/2$.  This implies that, inside a
cluster, the difference of internal phases between a pair of
elements can be larger than $\pi /2$, and their interaction can
thus be repulsive. However, the two elements can be part of the
same cluster due to the presence of other elements with
intermediate internal phases. These elements mediate the
interaction between the otherwise repulsive pairs, in such a way
that the effective force is attractive towards the cluster.

Once a cluster is formed, and after a certain transient has
elapsed, the total angle spanned by the distribution of internal
phases $\theta_i$ grows with time, due to the individual
evolution of each phase, Eq.~(\ref{eqtet}). This growth takes
place within time scales of order $\min_{ij} |\Omega_i-
\Omega_j|^{-1}$.  Under some simplifying assumptions, it can be
shown that a cluster becomes unstable when the total angle
spanned by the internal phases reaches $\pi$. Suppose that, at a
given moment, the internal phases of the $n$ elements belonging
to a cluster are uniformly distributed over the interval
$(0,\Theta)$. Suppose also that all the elements are at the same
point in space, except for a test element at a small distance
$\delta {\bf r}$ (with $|\delta {\bf r}|<r_0$) from the cluster.
If $n$ is sufficiently large, taking into account
Eqs.~(\ref{eqmot1}) to (\ref{vv}), the equation of motion of the
test element can be written as
\begin{eqnarray}
\delta \dot {\bf r} &=& -k\, \delta {\bf r} \sum_{i} \cos
(\theta_i -\theta_0) \approx k\, n \, \delta {\bf  r} \
\frac{1}{\Theta} \int_0^\Theta \cos
(\theta_0-\theta) d\theta = \nonumber \\
&=& k\, n\, \frac{\sin(\theta_0-\Theta)-\sin \theta_0 }{\Theta}\,
\delta{\bf r} ,
\label{test}
\end{eqnarray}
where $\theta_0$ is the internal phase of the test element. The
factor multiplying $\delta {\bf r}$ in the right-hand side of
Eq.~(\ref{test}) is negative for all $\theta_0 \in (0,\Theta)$ if
$0\le \Theta <\pi$. Under these conditions, the force on the test
element is attractive. As $\Theta$ reaches $\pi$, however, the
factor of $\delta {\bf r}$ becomes positive at $\theta_0=0$ and
$\pi$. This implies that, as soon as $\Theta >\pi$, the elements
with extreme internal phases, $\theta\approx 0$ or $\theta
\approx \pi$, will be repelled from the cluster. Due to their
mutual repulsion, elements with internal phases close to
$\theta=0$ and $\theta=\pi$ will move away from the cluster in
opposite directions. Moreover, as the distribution of phases in
the remaining of the cluster keeps broadening and new elements
reach the stability limits, they will be attracted by those
elements with similar phases which have just left the cluster,
and therefore move in their direction. This mechanism gives rise
to the ``strings'' of elements when a cluster becomes unstable
and disintegrates.

\begin{figure}
\centering
\resizebox{\columnwidth}{!}{\includegraphics[clip=]{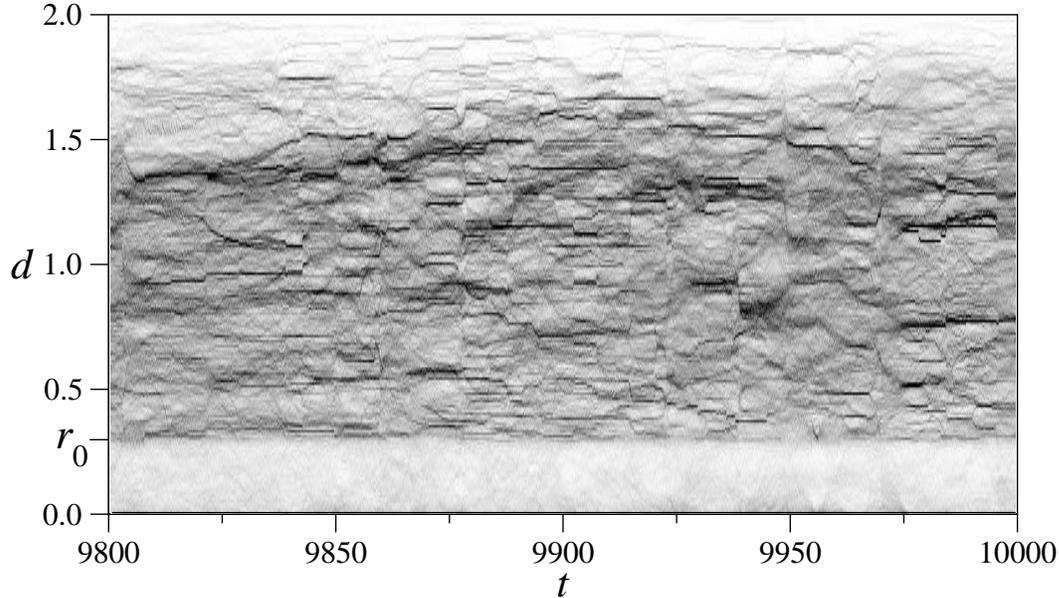}}
\caption{Density plot of the histogram of pair distances as a
function of time, for the same realization as in Fig.~\ref{figr2}
($N=100$, $r_0=0.3$, $\sigma=0.1$), along $200$ time units.
Darker tones correspond to higher concentrations. \label{figcl}}
\end{figure}

The dynamics of clustering is effectively illustrated by the
distribution of distances between elements \cite{zmr,zmn}.
Calculating the $N(N-1)/2$ Euclidean pair distances
\begin{equation}
d_{ij}=|{\bf r}_i - {\bf r}_j|
\end{equation}
for all $i$ and $j> i$, we construct a histogram  where the
height of the column with base $(d,d+\Delta d)$ is proportional
to the number of pair distances within that interval. Compact
clusters cause the appearance of sharp peaks in the histogram, at
$d\approx 0$ and other distances, while non-clustered elements
give rise to a smooth background. Figure \ref{figcl} shows, as a
density plot, the evolution of a $200$-column histogram of pair
distances $d\in (0,2)$ ($\Delta d=0.01$) for the same numerical
realization as in Fig.~\ref{figr2}, between $t=9800$ and $10000$.
As expected, the region $d<r_0$ is strongly depleted (except for
the peak at $d\approx 0$), because pairs of elements at such
distances either become entrained into clusters or are mutually
repelling. For $d>r_0$, dark horizontal lines reveal localized
compact clusters. As time elapses, these clusters can become less
well-defined, and they are seen to move, disperse, coalesce, or
split to form smaller groups.

This complex dynamics, driven by the autonomous evolution of the
internal phases, calls for a statistical description in terms of
a characterization of the partition of the ensemble into
clusters. With this aim, we introduce two order parameters
\cite{zmr}. The parameter $R$ is the time-averaged fraction of
pair distances smaller than a certain small threshold $\delta$.
It reads
\begin{equation}
R = \left\langle \frac{2}{N(N-1)} \sum_{i=1}^N
\sum_{j=i+1}^N H(\delta-d_{ij}) \right\rangle ,
\end{equation}
where $\langle \cdot \rangle$ indicates average over long times,
and $H(z)$ is the step function: $H(z)=1$ for $z>0$, and $H(z)=0$
otherwise. The second parameter, $S$, is given by the
time-averaged fraction of elements which have at least another
element as a distance smaller than $\delta$. It can be calculated
as
\begin{equation}
S = \left\langle 1- \frac{1}{N}
\sum_{i=1}^N \prod_{j=i+1}^N
H(\delta-d_{ij}) \right\rangle .
\end{equation}
Associating the threshold $\delta$ with the (maximal) size of a
cluster, both parameters are statistical measures of the degree
of clustering in the ensemble. Their values are close to zero if
most elements are not entrained, and reach unity if all elements
belong to clusters. They are, however, independent quantities.
For a fixed value of $S$, which measures the total population
inside clusters, $R$ is small if the number of clusters is large,
and {\it vice versa}. In other words, $R$ characterizes the
average number of elements per cluster. An approximate evaluation
of the number of clusters $M$ is obtained by assuming that all
clusters have the same population, as $M=S^2/R$ \cite{zmr}.

\begin{figure}
\centering
\resizebox{.8\columnwidth}{!}{\includegraphics[clip=]{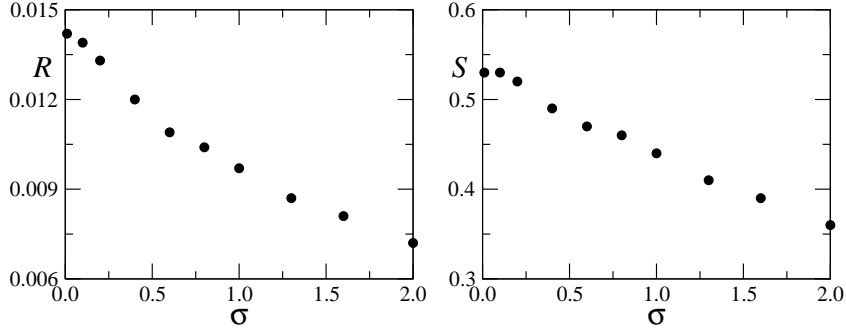}}
\caption{The order parameters $R$  and $S$ as functions of the
frequency dispersion $\sigma$, for an interaction range $r_0=0.3$.
\label{figr4}}
\end{figure}

To characterize clustering in our system, we have numerically
calculated the order parameters $R$ and $S$ as functions of the
frequency dispersion $\sigma$ and the interaction range $r_0$ in
a system of $N=100$ elements, taking $\delta=0.01$. Figure
\ref{figr4} shows $R$ and $S$ versus $\sigma$ for $r_0=0.3$. For
this intermediate value of the interaction range, the two order
parameters display a moderate decrease as the frequency
dispersion grows. As frequency differences become larger, the
typical length of the time intervals during which the interaction
of a given pair of elements is attractive shortens. Consequently,
shorter times are available for the aggregation of elements into
clusters and, thus, the average degree of clustering diminishes.
The approximate number of clusters, calculated as $M=S^2/R$, is
practically constant, $M\approx 20$, in the considered interval
of variation for $\sigma$. The same behaviour is found for
smaller values of the interaction range. On the other hand, for
large values of $r_0$, the dependence with the frequency
dispersion becomes less trivial. We analyze this specific
situation in the next section.

\begin{figure}
\centering
\resizebox{.8\columnwidth}{!}{\includegraphics[clip=]{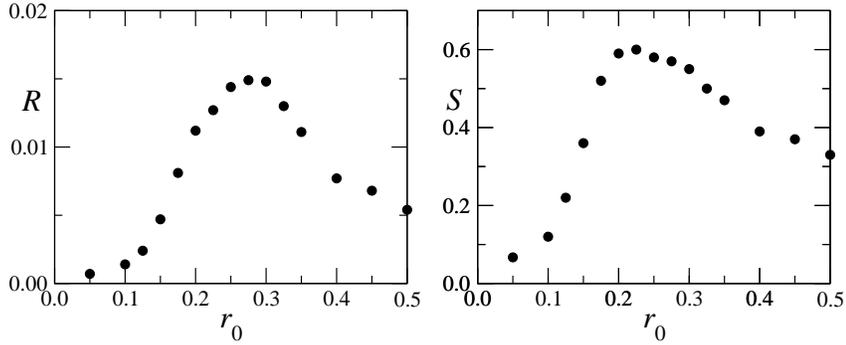}}
\caption{The order parameters $R$ and $S$ as functions of the
interaction range $r_0$, for a frequency dispersion $\sigma=0.1$.
\label{figr3}}
\end{figure}

The dependence of the clustering order parameters $R$ and $S$ on
the interaction range $r_0$ for fixed $\sigma$ is more
interesting, as shown in Fig.~\ref{figr3} for $\sigma=0.1$. We
find that both parameters display a maximum, at $r_0\approx 0.27$
for $R$ and $r_0\approx 0.23$ for $S$.  For smaller values of the
interaction range, the two parameters rapidly drop to zero, while
for larger $r_0$ they exhibit a smoother decay. The approximate
number of clusters,  $M=S^2/R$, grows steadily from $M \approx 1$
for $r_0\approx 0$ and attains a sharp  maximum, $M\approx 30$,
at $r_0\approx 0.25$. For larger interaction ranges, it soon
reaches an almost constant value, $M\approx 20$.

\begin{figure}
\centering
\resizebox{.7\columnwidth}{!}{\includegraphics[clip=]{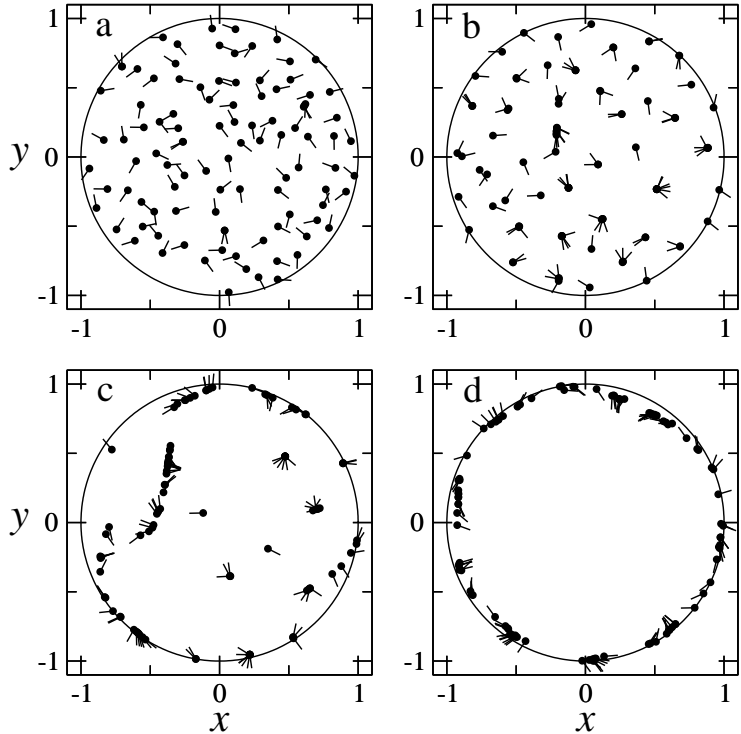}}
\caption{Four snapshots of an ensemble of size $N=100$ at times
$t=10000$, with $\sigma=0.1$ and different interaction ranges:
(a) $r_0=0.1$, (b) $r_0=0.2$, (c) $r_0=0.3$, and (d) $r_0=0.5$.
\label{figr1}}
\end{figure}

In order to explain the non-monotonic dependence of the
clustering order parameters with the interaction range, it is
useful to inspect the instantaneous spatial distribution of the
ensemble for different values of $r_0$ (Fig.~\ref{figr1}). For
sufficiently small $r_0$, there is little chance that two
elements are simultaneously found within the interaction range.
In our two-dimensional system, the average distance $\bar d$
between elements can be evaluated as $\bar d =\sqrt{A/N}$, where
$A$ is the area of the domain accessible to the ensemble. Taking
$A=\pi r_{\max}^2$ and $r_{\max} =1$, we find $\bar d \approx
0.18$. Figures \ref{figr1}a and b show that for $r_0<\bar d$ only
a few elements form clusters, whereas for $ r_0 \gtrsim \bar d$
most of the ensemble is clustered in many groups. For such values
of the interaction range, consequently, the clustering order
parameters increase with $r_0$ (Fig.~\ref{figr3}).

As the interaction range grows, the collective organization of
elements in larger spatial structures becomes possible. In
particular, the ``strings'' associated with  unstable clusters,
discussed above, entrain now a substantial part of the ensemble
(Fig.~\ref{figr1}c). The contribution of these structures to the
clustering order parameters is expected to be considerably weaker
than that of compact clusters, which explains the decline of $R$
and $S$ for $r_0 \gtrsim 0.3$. For even larger interaction ranges
the dynamics is dominated by the remnant repulsive interaction
between compact clusters. When $r_0$ is sufficiently large, all
the elements are found most of the time at the boundary of the
spatial domain (Fig.~\ref{figr1}d). Note that  elements are still
organized in space according to their internal state: generally,
neighbouring elements have similar internal phases. This form of
organization in the limit of large interaction ranges is treated
in detail in the next section.

\begin{figure}
\centering
\resizebox{.7\columnwidth}{!}{\includegraphics[clip=]{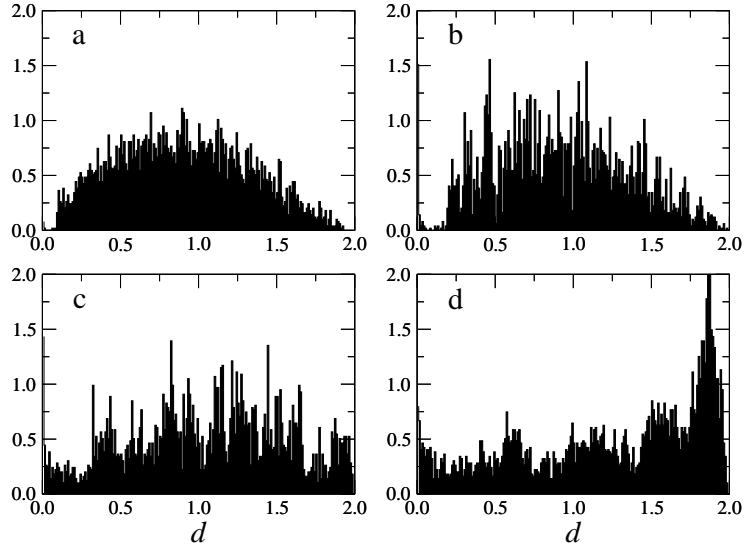}}
\caption{Normalized histograms of pair distances for the
snapshots of Fig.~\ref{figr1}. (a) $r_0=0.1$, (b) $r_0=0.2$, (c)
$r_0=0.3$, and (d) $r_0=0.5$. \label{figr1a}}
\end{figure}

The change in the collective spatial distribution as the
interaction range grows is also apparent in the distribution of
pair distances $d_{ij}$. Figure \ref{figr1a} shows $200$-column
normalized histograms of pair distances for the snapshots shown
in Fig.~\ref{figr1}. When the degree of clustering is small
(Fig.~\ref{figr1a}a) the distribution of pair distances is rather
smooth, and no significant contributions are present at $d\approx
0$. As the interaction range grows and clustering becomes more
important, the distribution spreads and develops sharp spikes.
Note, in particular, the high peak at $d\approx 0$
(Fig.~\ref{figr1a}b and c). For even larger $r_0$ the degree of
clustering recedes, the peak at $d\approx 0$ becomes smaller, and
other spikes are replaced by broader structures
(Fig.~\ref{figr1a}d). The highest peak, centered at $d\approx
1.8$, is a direct byproduct of the concentration of elements in
the boundary of the spatial domain.

The fact that, in the limit of large interaction range, the
ensemble is mainly concentrated in the boundary of the accessible
domain suggest a reduced representation of the problem, where the
position of each element is given by a linear coordinate along
the boundary. In the case of the circular domain considered
above, such approximation yields a set of dynamical equations
which are formally equivalent to the model equations of coupled
phase oscillators. As shown in the next section, this reduced
representation admits to be studied analytically to a
considerable extent.

\section{Phase oscillators with time-dependent coupling}
\label{sect2}

\subsection{Reduction to phase variables}

\begin{figure}
\centering
\resizebox{.5\columnwidth}{!}{\includegraphics[clip=]{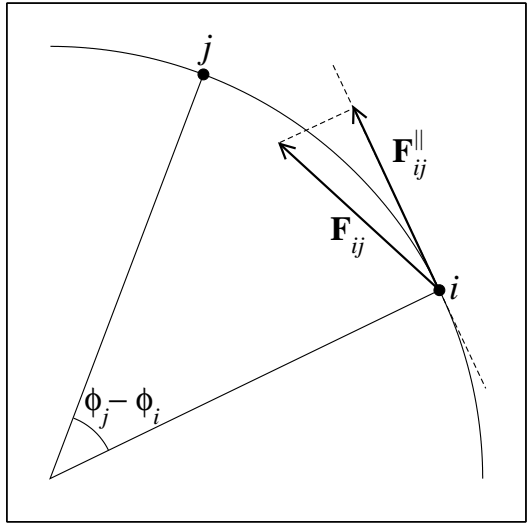}}
\caption{Reduction of the circular-domain problem to phase
variables, in the limit of large interaction range. \label{fana}}
\end{figure}

When the range of interactions is larger than the linear size of
the system, all the elements are subject to the action of the
whole ensemble at all times. As discussed in the preceding
section, the regime of large $r_0$ is characterized by the
repelling interaction of clusters, which drives the elements to
the boundaries of the accessible spatial domain. For the case of
the circular domain considered above, the position of an element
at the boundary can be characterized by the angle $\phi_i$
defined by the element and an arbitrarily chosen origin, with
vertex at the center of the circle. Assuming that the elements do
not abandon the boundary at any time, the evolution of $\phi_i$ is
governed by
\begin{equation}
\dot \phi_i =-\dot x_i \sin \phi_i+\dot y_i \cos \phi_i =
\sum_{j\neq i} V(\theta_i,\theta_j) F_{ij}^\parallel ,
\end{equation}
where $F_{ij}^\parallel$ is the projection of the pair force
${\bf F} ({\bf r}_i-{\bf r}_j)$ parallel the circular boundary
[cf.~Eq.~(\ref{eqmot1})]. Figure \ref{fana} shows the geometry of
the problem. With $r_0>2r_{\max} =2$, the modulus of the force
(\ref{ff}) between any pair of elements is given by
\begin{equation}
F_{ij} = k\, d_{ij} =k\sqrt{2[1-\cos(\phi_j-\phi_i)]}=2k \sin \left(
\frac{\phi_j-\phi_i}{2} \right) .
\end{equation}
The angle between the force and the tangent to the circle at the
position of element $i$ is $(\phi_j-\phi_i)/2$. Therefore,
\begin{equation}
F_{ij}^\parallel = F_{ij} \cos \left( \frac{\phi_j-\phi_i}{2} \right)
= k \sin(\phi_j-\phi_i) .
\end{equation}

The equation of motion for $\phi_i$ becomes
\begin{equation} \label{eqmot2}
\dot \phi_i =
k \sum_{j\neq i} V(\theta_i,\theta_j)  \sin (\phi_j-\phi_i).
\end{equation}
Remarkably, this equation is formally equivalent to the model
equation for an ensemble of phase oscillators with identical
natural frequencies $\omega=0$, where the state of each oscillator
is characterized by its phase $\phi_i$. The collective behaviour
of ensembles of phase oscillators has been thoroughly analyzed
during the last two decades. With $V(\theta_i,\theta_j) \equiv 1$
for all $i,j$, Eq.~(\ref{eqmot2}) reduces to the Kuramoto model
for globally coupled identical oscillators \cite{kura}. In this
case, full synchronization ($\phi_i(t)=\phi_j(t)$ for all $i,j$)
is a stable asymptotic state for any positive value of $k$.
Time-independent interaction weights, where $V(\theta_i,\theta_j)
\equiv V_{ij}$ is constant but can have different values for each
oscillator pair,  have been shown to induce typical features of
disordered systems, such as glassy-like behaviour, frustration,
and algebraic relaxation towards equilibrium
\cite{d622,d1073,d2145}. As we show in the following, when the
interaction between phase oscillators is modulated by the
time-dependent interaction weight $V(\theta_i,\theta_j)$ defined
by Eqs.~(\ref{eqtet}) and (\ref{vv}), various forms of collective
evolution --including dynamical clustering-- emerge.  The
transition between these regimes is controlled by $\sigma$, the
frequency dispersion of the internal phases $\theta_i$, which is
now the only parameter left in the model.

\subsection{Order-disorder critical transition} \label{sect32}

\begin{figure}
\centering
\resizebox{\columnwidth}{!}{\includegraphics[clip=]{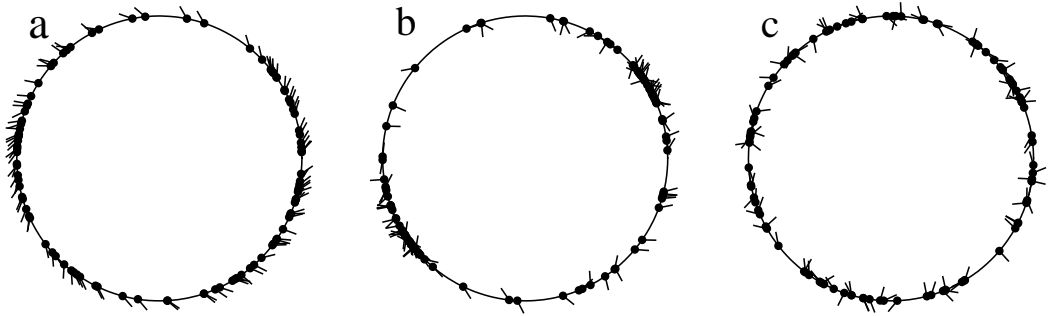}}
\caption{Long-time snapshots of an ensemble of $100$ phase
oscillators, for different values of the frequency dispersion:
(a) $\sigma=0.1$, (b) $\sigma=0.3$, and (c) $\sigma =0.5$.
\label{figo1}}
\end{figure}

We begin our analysis of Eq.~(\ref{eqmot2}) by performing
numerical simulations of the system for ensembles ranging from
$N=10^2$ to $10^6$. As customary with coupled phase oscillators,
we rescale the constant $k$ with the ensemble size as $k=K/N$,
and fix $K=1$. The frequencies $\Omega_i$ of the internal phases
are drawn from the distribution of Eq.~(\ref{Gaussian}). We fix
$\Omega_0=0$, and consider various values of the frequency
dispersion $\sigma$. Figure \ref{figo1} shows snapshots of the
ensemble taken at sufficiently long times, for three values of
$\sigma$. For small $\sigma$ (Fig.~\ref{figo1}a), the phases
$\phi_i$ are homogeneously distributed around the circle, and a
high degree of correlation between $\phi_i$ and the internal
variables $\theta_i$ is apparent. For large frequency dispersion,
the distribution of phases is also approximately homogeneous but,
on the other hand, the correlation with the internal variables is
lost (Fig.~\ref{figo1}c).  At intermediate values of $\sigma$, we
find strongly heterogeneous distributions for the phases
$\phi_i$. Figure \ref{figo1}b  shows a state where most of the
ensemble is split into two clusters with opposite phases.

To explain the appearance of these three regimes we first note
that, for small $\sigma$, the internal variables --and,
therefore, the interactions-- evolve slowly. Over such long time
scales, the phases $\phi_i$ are able to follow, almost
adiabatically, the evolution of the interaction weights
$V(\theta_i,\theta_j)$. Numerical results show that, in this
situation, the phase $\phi_i$ and the internal variable
$\theta_i$ of each oscillator are related according to
\begin{equation} \label{phitet}
\phi_i = \pm \theta_i +\phi_0.
\end{equation}
The sign factor of $\theta_i$ and the value of $\phi_0$ are the
same for all the oscillators in the ensemble. They are determined
by the initial condition, and $\phi_0$ may slowly change with
time. In the limits $\sigma\to 0$ and $N\to \infty$, a
homogeneous distribution of phases related to the internal
variables as in Eq.~(\ref{phitet}) is a solution to
Eq.~(\ref{eqmot2}). In fact, if such relation holds, we have
\begin{eqnarray}
\dot \theta_i = \mp \dot \phi_i &=& \pm \frac{1}{N} \sum_j \cos
(\theta_i-\theta_j) \sin (\theta_i-\theta_j) \nonumber \\
&\approx& \pm  \frac{1}{2\pi} \int_0^{2\pi} \cos
(\theta_i-\theta)\sin (\theta_i-\theta) d\theta =0.
\end{eqnarray}
For finite values of $\sigma$, the relation between $\phi_i$ and
$\theta_i$ is maintained as long as the frequency dispersion
remains small. For large $\sigma$, on the other hand, most
interaction weights show large changes over short times, and the
phases are not able to adjust their values to such changes. In
this situation, each oscillator is subject to rapidly fluctuating
forces and, as a result, phases are homogeneously distributed,
but no relation with the internal variables can persist.

\begin{figure}
\centering
\resizebox{.8\columnwidth}{!}{\includegraphics[clip=]{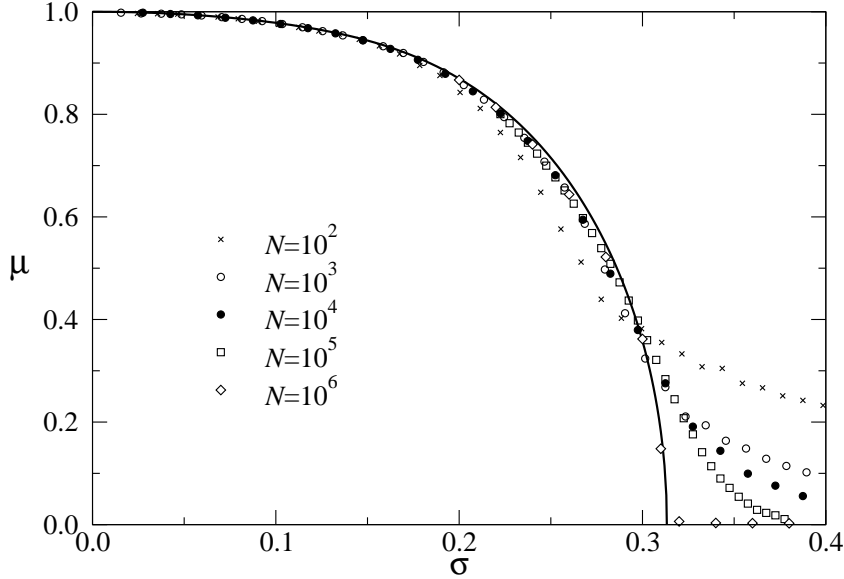}}
\caption{The order parameter $\mu$ as a function of the frequency
dispersion $\sigma$, for ensembles of various sizes $N$. The
curve corresponds to the analytical evaluation of $\mu$, from
Eq.~(\ref{mu}). \label{clusaa}}
\end{figure}

The transition between the ordered and the disordered regime,
respectively found for small and large frequency dispersion, can
be characterized by means of a suitably defined order parameter.
With this aim, we first introduce for each oscillator the
two-dimensional complex vector ${\bf m}_i =(\exp[{\rm
i}(\phi_i-\theta_i)],\exp[{\rm i}(\phi_i+\theta_i)])$, and define
the average
\begin{equation} \label{bfm}
{\bf m} = \frac{1}{N} \sum_i {\bf m}_i
\equiv (\mu_+ \exp({\rm i} \psi_+), \mu_- \exp({\rm i} \psi_-)) .
\end{equation}
The average of the modulus $\mu = |{\bf m}| =
\sqrt{\mu_+^2+\mu_-^2}$ over sufficiently long times is an
adequate order parameter. Indeed, if relation (\ref{phitet}) is
approximately verified, we have $\mu \approx 1$. If, on the other
hand, the values of $\phi_i$ and $\theta_i$ are uncorrelated, we
find $\mu \sim 1/\sqrt{N}$. Figure \ref{clusaa} shows the
numerical evaluation of $\mu$ as a function of the frequency
dispersion $\sigma$, for different values of $N$. We note that,
as $N$ grows, the order parameter develops an abrupt inflection
at $\sigma \approx 0.3$, which suggests the presence of a critical
phenomenon. As we show in the following, this critical phenomenon
can be described analytically if a few assumptions, supported by
numerical results, are made.

Taking into account that, in the limit $\sigma \to 0$, the
relation (\ref{phitet}) is equivalent to the stationarity
condition for an ensemble with homogeneous phase distribution,
for larger values of $\sigma$ we can write
\begin{equation} \label{delta}
\phi_i=\pm (\theta_i-\delta_i)+\phi_0,
\end{equation}
where $\delta_i(t)$ measures the deviation of each oscillator
with respect to the stationary state of small frequency
dispersion. We assume that, for sufficiently large ensembles and
long times, the distribution of these deviations is given by a
well-defined function $p(\delta)$, satisfying
$p(\delta)=p(-\delta)$. Implicitly, this also amounts to
conjecture that the values of $\delta_i$ on the one hand, and
$\phi_i$ and $\theta_i$ on the other, are not correlated.
Replacing Eq.~(\ref{delta}) into the definition of ${\bf m}_i$
and using Eq.~(\ref{bfm}) to calculate the order parameter $\mu$,
we find
\begin{equation} \label{mu0}
\mu = \int_{-\pi}^{\pi} p(\delta) \cos \delta d\delta.
\end{equation}

Assuming that $\phi_0$ in Eq.~(\ref{delta}) is independent of
time, Eqs.~(\ref{eqmot2}) and (\ref{mu0}) yield the following
evolution equation for $\delta_i$:
\begin{equation}
\dot \delta_i = \Omega_i -\frac{\mu}{2} \sin \delta_i .
\end{equation}
This equation has been extensively discussed in connection with
the dynamics of globally coupled phase oscillators \cite{saka}.
For $|\Omega_i|<\mu /2$, $\delta_i$ has two fixed points, at the
solutions of
\begin{equation} \label{deltaeq}
\sin \delta_i =\frac{2\Omega_i }{\mu} .
\end{equation}
One of them is stable, and the other is unstable. On the other
hand, for $|\Omega_i|>\mu /2$, there are no stable fixed points
and $|\delta_i|$ grows indefinitely with time.  We can therefore
distinguish between two subpopulations in the ensemble, with
qualitatively different behaviour. Those oscillators for which
$|\Omega_i|<\mu /2$ reach, at asymptotically large times, a
stationary deviation $\delta_i$. In contrast, for those
oscillator with $|\Omega_i|>\mu /2$ the phase $\phi_i$ does not
reach a stationary value with respect to the internal variable
$\theta_i$. We refer to these two subpopulations as subensembles
I and II, respectively. Whether a given oscillator belongs to one
of the two subensembles depends on the value of $\mu$. According
to Eq.~(\ref{mu0}), $\mu$ is given by the distribution of
stationary deviations $\delta_i$, which in turn depend on $\mu$
through Eq.~(\ref{deltaeq}).  Thus, the order parameter must be
found self-consistently.

The distribution of deviations $p(\delta )$ consists of
contributions from each sub\-ensemble $p = p_{\rm I}+p_{\rm II}$.
Since the deviations $\delta_i$  for the oscillators of
subensemble II are not stationary and move at different
velocities, $p_{\rm II} (\delta)$ is a flat distribution and,
thus, does not contribute to $\mu$ in Eq.~(\ref{mu0}). The
contribution of subensemble I, on the other hand,  can be found
from the identity $p_{\rm I} (\delta) d\delta =g(\Omega)
d\Omega$, taking into account the connection between the deviation
and the internal frequency of each oscillator, given through
$\mu$ by Eq.~(\ref{deltaeq}). Replacing $p(\delta)$ into
Eq.~(\ref{mu0}) yields
\begin{equation} \label{mu}
\mu = \int_{-\mu/2}^{\mu/2} g(\Omega)
\sqrt{1-\frac{4\Omega^2}{\mu^2}} d\Omega.
\end{equation}
This is the self-consistency equation which makes it possible to
find $\mu$, given the distribution of internal frequencies
$g(\Omega)$. For our Gaussian distribution, Eq.~(\ref{mu}) has a
nontrivial solution $\mu>0$  if $\sigma<\sigma_c = \sqrt{\pi /32}
\approx 0.313$.  As the frequency dispersion approaches the
critical value $\sigma_c$ from below, the order parameter
vanishes as $\mu \sim |\sigma-\sigma_c|^{1/2}$. For
$\sigma>\sigma_c$, the only solution is $\mu=0$. The curve in
Fig.~\ref{clusaa} shows the solution to Eq.~(\ref{mu}) as a
function of the frequency dispersion. The excellent agreement
with numerical results validate our analytical approach. Remnant
discrepancies can be ascribed to the enhancement of finite-size
effects around the critical transition.

Note that the same singular dependence of the order parameter
$\mu$ close to the order-disorder transition would be found for a
broad class of frequency distributions $g(\Omega)$. In fact,
since only small values of $\Omega$ contribute to the integral in
Eq.~(\ref{mu}) when $\mu$ is small, the critical behaviour is
determined by the shape of the maximum of $g(\Omega)$ at
$\Omega=0$. Suppose that, for $\Omega \approx 0$, we have
$g(\Omega) \approx g(0) -\gamma \Omega^2$. Replacing into
Eq.~(\ref{mu}) we find
\begin{equation}
\mu = 8 \sqrt{\frac{1}{\pi \gamma} \left[1-\frac{\pi}{4} g(0) \right]}.
\end{equation}
Consequently, the critical point is determined by the equation
$g(0) = 4/\pi$. If, sufficiently close to the point where this
equation holds, $g(0)$ varies linearly with the control parameter
$\sigma$, the order parameter will behave as $\mu \sim
|\sigma-\sigma_c|^{1/2}$. The critical exponent, here equal to
$1/2$, changes if the maximum of $g(\Omega)$ at $\Omega=0$ is not
quadratic.

Using the complex vector $\bf m$ defined in Eq.~(\ref{bfm}), the
equation of motion (\ref{eqmot2}) for the phase of an individual
oscillator can be recast as
\begin{equation} \label{mupm}
\dot \phi_i =\frac{\mu_+}{2} \sin(\psi_+ +\theta_i-\phi_i) +
\frac{\mu_-}{2} \sin(\psi_- -\theta_i-\phi_i) .
\end{equation}
The form of this equation emphasizes the mean-field nature of the
interactions, as $\dot \phi_i$ depends only on the individual
variables $\phi_i$ and $\theta_i$ and on the averages involved in
the definition of $\bf m$. Since for asymptotically large
ensembles, $N\to \infty$, $\mu=\sqrt{\mu_+^2+\mu_-^2}$ is zero
for $\sigma>\sigma_c$,  both $\mu_+$ and $\mu_-$ must also vanish
beyond the transition. This implies that in the thermodynamical
limit and for $\sigma>\sigma_c$, $\dot \phi_i=0$ and the dynamics
is thus frozen. For finite systems, $\mu_+$ and $\mu_-$ are of
order $N^{-1/2}$. The evolution of individual phases can
therefore be thought as driven by random forces of that order.
Below the transition, evolution becomes progressively slower as
$\sigma$ approaches the critical point.

\subsection{Dynamical clustering}

\begin{figure}
\centering
\resizebox{\columnwidth}{!}{\includegraphics[clip=]{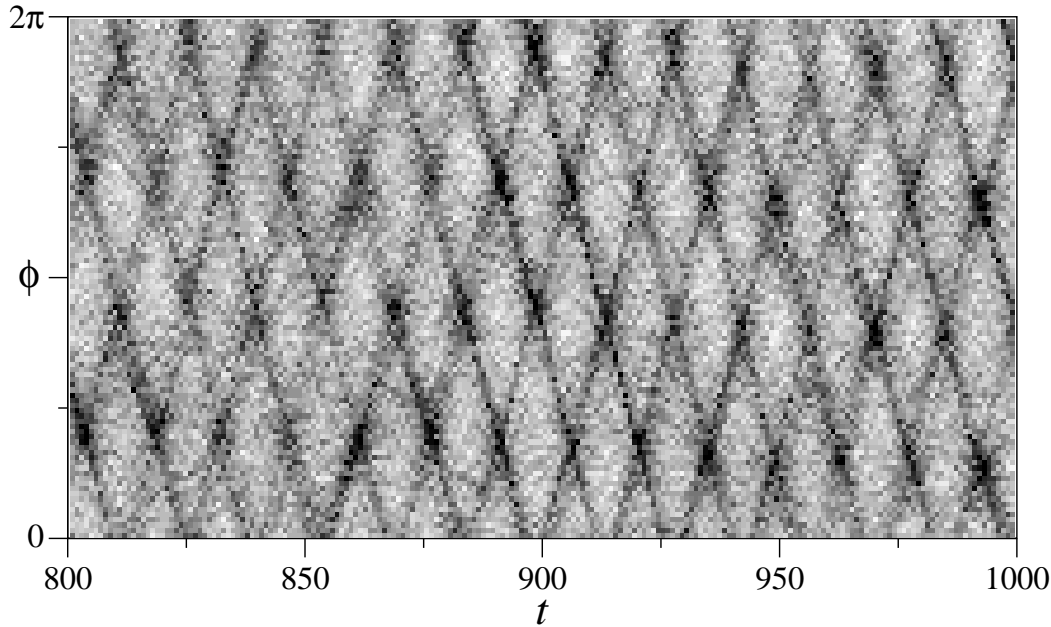}}
\caption{Density plot of the phase distribution ($100$-column
histogram) as a function of time, for a system of $N=10^3$
oscillators with  frequency dispersion $\sigma=0.3$, along $200$
time units. Darker tones correspond to higher concentrations.
\label{figo3}}
\end{figure}

As pointed out above, the dynamical regimes of small and large
frequency dispersion are separated by a region, around
$\sigma\approx 0.3$,  where clustering is found
(Fig.~\ref{figo1}b). To disclose the properties of this
intermediate regime, we have numerically studied the evolution of
the distribution of phases $\phi_i$. Recording the evolution of
each phase, we construct a histogram where the height of  the
column with base $(\phi,\phi+\Delta \phi)$ is proportional to the
number of oscillators whose phases lie within that interval.
Figure \ref{figo3} shows a density plot of such histogram as a
function of time, for $\sigma=0.3$.

\begin{figure}
\centering
\resizebox{.8\columnwidth}{!}{\includegraphics[clip=]{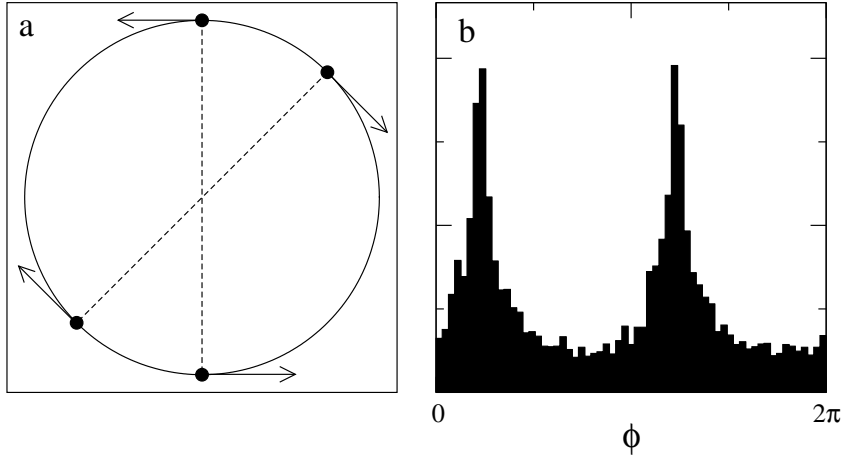}}
\caption{(a) Schematic representation of the relative position
and motion of the two pair of clusters found in the clustering
regime. (b) Histogram of the distribution of phases when the two
cluster pairs cross each other, in a system of $N=10^3$
oscillators with frequency dispersion $\sigma=0.3$. \label{clus0}}
\end{figure}

We note that a fraction of the ensemble is entrained into four
clusters. A close inspection of these clusters shows that they
are formed by the oscillators of subensemble II (see
Sect.~\ref{sect32}) with internal frequencies just above their
lowest value, $|\Omega_i| \gtrsim \mu/2$. Though the clusters are
not sharply localized, it is still possible to define their phase
by averaging $\phi_i$ over the oscillators belonging to each
cluster. Two of these clusters, whose phases differ by $\pi$,
move at a well defined velocity in one direction. The other two,
whose phases also differ by $\pi$, move in the opposite direction
and with the same speed. The two pairs of clusters are formed by
oscillators whose internal frequencies have opposite signs.
Figure \ref{clus0}a illustrates the situation. Due to their
contrary motion, the two pairs of clusters cross each other
recurrently.  When these crossings happen, their populations are
temporarily superimposed in phase. Consequently, two big clusters
with opposite phases are recurrently built up (Figs.~\ref{figo3}
and \ref{clus0}b). Figure \ref{figo1}b shows a $100$-oscillator
ensemble at one of such events.

\begin{figure}
\centering
\resizebox{.8\columnwidth}{!}{\includegraphics[clip=]{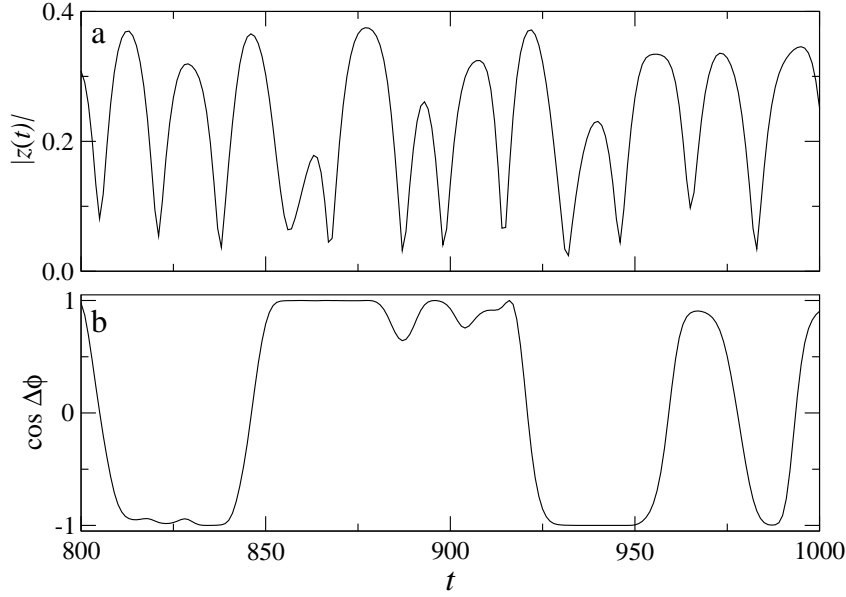}}
\caption{(a) Evolution of $|z(t)|$, Eq.~(\ref{z}), for a system
of $N=10^3$ oscillators wit h frequency dispersion $\sigma=0.3$.
(b) Phase difference of two oscillators of subensemble II with
similar frequencies, in the same realization as in (a).
\label{clus00}}
\end{figure}

The recursive formation of the two anti-phase big clusters can be
quantitatively characterized by an order parameter. The modulus
of the complex quantity
\begin{equation} \label{z}
z(t) =\frac{1}{N} \sum_i \exp (2{\rm i} \phi_i)
\end{equation}
is approximately equal to one if the whole ensemble splits into
two anti-phase well-localized clusters of similar sizes, while
for incoherent or higher-order clustered states we have $|z| \sim
N^{-1/2}$. Figure \ref{clus00}a shows the evolution of $|z(t)|$
in the clustering regime, displaying its irregular oscillations
between small and large values.  For the same realization,
Fig.~\ref{clus00}b shows the phase difference $\Delta\phi=
\phi_a-\phi_b$ for two oscillators $a$ and $b$ in subensemble II.
Large positive values of $\cos \Delta \phi$ identify states where
the two oscillators are part of the same cluster, while negative
values correspond to the situation  where they are in different
clusters. The intermittent transitions of $\cos \Delta\phi$
between positive and negative values demonstrate the dynamical
nature of clustering, where clusters maintain their identity but
elements can migrate between them.

\begin{figure}
\centering
\resizebox{.8\columnwidth}{!}{\includegraphics[clip=]{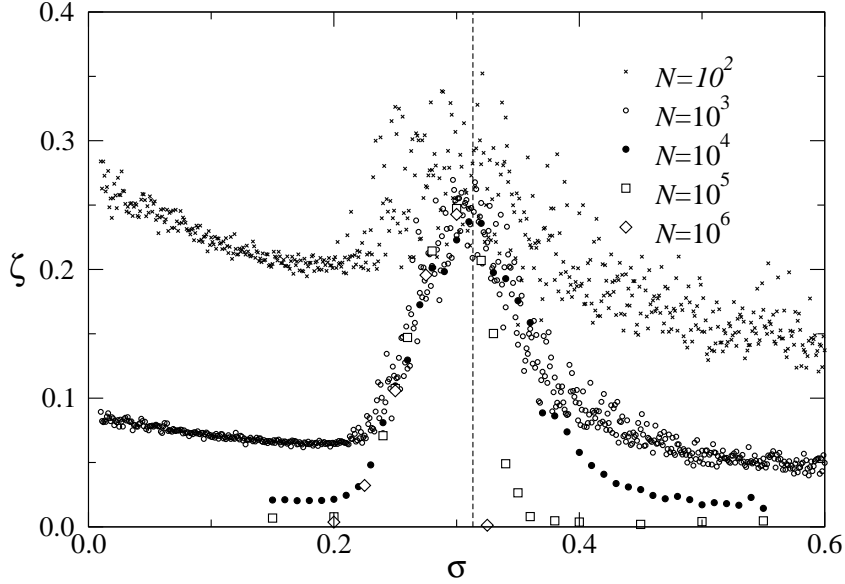}}
\caption{The order parameter $\zeta$ as a function of the
frequency dispersion $\sigma$ for oscillator ensembles of various
sizes. The vertical dashed line indicates the critical point of
the order-disorder transition, $\sigma_c = \sqrt{\pi/32}$.
\label{clusb}}
\end{figure}

The average of $|z(t)|$ over long times provides a suitable order
parameter, $\zeta$, for the detection of clustered states. Note
that $\zeta$ and $\mu$ are independent order parameters, as they
measure  Fourier contributions of different order to the
distribution of phases. Figure \ref{clusb} shows numerical
results for $\zeta$ as a function of the frequency dispersion. As
$\sigma$ approaches the critical value $\sigma_c$ from below,
$\zeta$ grows and exhibits a peak just at the left of the
critical point. The profile of this peak becomes better defined
as the ensemble size $N$ grows. The order parameter grows from
zero at $\sigma \approx 0.2$ and reaches the maximum value $\zeta
\approx 0.25$ for $\sigma \lesssim \sigma_c$. Beyond the
transition, $\zeta$ drops abruptly. Results for the largest
ensembles suggest that, in the thermodynamical limit, it vanishes
for $\sigma>\sigma_c$. The suppression of clustering beyond the
transition is a consequence of the fact that, as discussed in
Sect.~\ref{sect32}, the evolution of asymptotically large
ensembles becomes frozen. As $\sigma$ approaches the critical
point and the order parameter $\mu$ tends to vanish, the typical
evolution time scales --which, according to Eq.~(\ref{mupm}), are
of order $\mu^{-1}$-- are increasingly large. Thus, the formation
of clusters is progressively retarded and, finally, it is
suppressed.

The appearance of the four clusters in the transition regime can
be qualitatively understood taking into account that, for an
oscillator with internal frequency $\Omega_i$, the most
persistent interactions affecting its dynamics are due to
oscillators with similar frequencies, $\Omega_j \approx
\Omega_i$. In such case, in fact, the interaction weight
$V(\theta_i, \theta_j)$ remains almost constant over very long
time scales, of order $|\Omega_i - \Omega_j|^{-1}$. Fixing the
attention on a group of oscillators with similar frequencies, we
realize that their interactions are weighted by almost
time-independent factors which, however, can take a different
value for each oscillator pair. We can have both positive and
negative interaction weights. Now, it is known that an ensemble
of coupled oscillators with disordered interactions forms, if the
level of frustration is moderate, two anti-phase localized
clusters \cite{d622}. In our system, triggering the formation of
these two clusters requires moreover that the phases $\phi_i$ are
not pinned to the internal variables $\theta_i$, and that a
sufficiently large population with similar frequencies actually
exist. These two conditions are met by the oscillators of
subensemble II with frequencies close to the limit $|\Omega|=
\mu/2$, where the density is larger. We expect that two clusters
form in the region of positive frequencies, and other two
clusters appear for $\Omega<0$. The symmetry of this
configuration insures that the two pairs of clusters will move in
opposite direction and with the same speed. As far as $\sigma$
remains below the transition point, increasing the frequency
dispersion contributes in two cooperating ways to the growth of
the population of these clusters: larger values of $\sigma$ imply
more oscillators with higher internal frequencies and, at the
same time, a larger fraction in subensemble II, as $\mu$
decreases.

\section{Conclusion}
\label{sect3}

We have studied, both numerically and analytically, the dynamical
regimes of an ensemble of coupled motile elements whose
interactions are determined by the evolution of their internal
states. The interaction of each pair of elements is alternatively
attractive or repulsive, changing its sign within time scales
associated with the relative variation of the respective internal
variables. Though in our study we have considered specific forms
both for the spatial dependence of the interactions and for their
modulation by the internal state, our main results are expected
to hold for a broad class of similar models. The following
summary is focused on such generic conclusions.

Transitions between different dynamical regimes are controlled by
two parameters: the range of interactions and the typical
evolution time of the internal variables. The latter determines,
in turn, the time scales within which interactions change their
sign. The simultaneous presence of both positive and negative
pair interactions induces the formation of localized clusters of
mutually attracting elements. The remnant repulsion drives
clusters away from each other, a tendency to expansion that must
be counteracted by suitable boundary conditions. Due to the
change of the sign of interactions, clustering is dynamical, as
elements intermittently aggregate, disperse, and migrate between
clusters. The actual development of localized clusters, however,
requires that the interaction range is neither too small nor too
large. Clusters are practically not formed if that range is
smaller than the average distance between elements, while for
large interaction ranges other kinds of spatial organization,
such as ``strings'' of elements and condensation in the
boundaries, become more frequent.

When the range of interactions is large as compared with the
spatial size of the system, interactions act in conditions similar
to those of global coupling \cite{kura}, though different
interaction weights are still possible between different element
pairs. In such conditions, the spatial distribution of the
ensemble becomes less relevant to the collective dynamics, because
interactions are now independent of the relative position of
elements. Yet, their individual positions are still determined by
their interactions, which are now mainly driven by the internal
variables. When the evolution of the internal variables is
sufficiently slow, the spatial distribution can follow it almost
adiabatically, and the relative position of the elements adjusts
to their instantaneous internal state. As a result, a state of
high correlation between spatial and internal variables develops.
Due to the autonomous internal evolution of each element, this
ordered state is dynamical, and the spatial distribution keeps
varying with time. On the other hand, if the internal variables
change too fast, the spatial coordinates fail to follow their
evolution, and correlation between them breaks down. If the
internal evolution is sufficiently fast and the time average of
pair interactions is zero, the spatial distribution of the
ensemble becomes frozen. In our specific model, the transition
between the ordered and the disordered state has the properties
of a critical phenomenon. In the disordered state, evolution is
frozen in the limit of an infinitely large ensemble. The
transition is mediated by a regime of dynamical clustering,
induced by the persistent interaction between elements whose
internal variables change within similar time scales, and whose
positions and internal states are not strongly correlated.

The dynamics of internal states is expected to be an essential
ingredient in the description of systems whose collective
evolution resembles the complex collective dynamics of biological
and social populations \cite{axel}. Spontaneous development of
coherent behaviour may be enhanced if individual internal
variables are allowed to interact with each other. In fact, the
organization of the ensemble at the level of the individual
internal states, due to their mutual interaction, would imply
coherence also at the level of other variables, such as the
spatial position. These processes are expected to be relevant to
the modeling of social-like mechanisms, of the type of imitation,
social influence, and formation of coalitions and hierarchical
structures. The investigation of their role in the emergence of
order has already been initiated in the literature \cite{shk},
also in connection with evolutionary adaptive phenomena
\cite{nuevo,ik1,ik2}.

\end{document}